\documentclass[manuscript,screen]{acmart}
\AtBeginDocument{%
  }

\author{Atrisha Sarkar}
\affiliation{%
   \institution{Western University}
   \city{London}
   \state{Ontario}
   \country{Canada}}
\email{atrisha.sarkar@uwo.ca}

\author{Isam Faik}
\affiliation{%
    \institution{Ivey Business School}
   \institution{Western University}
   \city{London}
   \state{Ontario}
   \country{Canada}}
\email{ifaik@ivey.ca}

\usepackage{booktabs}
\usepackage{float}
\usepackage{multirow}
\usepackage{etoolbox}
\AtBeginEnvironment{quote}{\par\small}
\renewenvironment{quote}{%
  \list{}{%
    \leftmargin0.5cm   
    \rightmargin\leftmargin
  }
  \item\relax
}
{\endlist}

\begin{document}

\title{Structural transparency of societal AI alignment through Institutional Logics.}





\begin{abstract}
The field of AI alignment is increasingly concerned with the questions of how values are integrated into the design of generative AI systems and how their integration shapes the social consequences of AI. However, existing transparency frameworks focus on the informational aspects of AI models, data, and procedures, while the institutional and organizational forces that shape alignment decisions and their downstream effects remain underexamined in both research and practice. To address this gap, we develop a framework of \emph{structural transparency} for analyzing organizational and institutional decisions concerning AI alignment, drawing on the theoretical lens of Institutional Logics. We develop a categorization of organizational decisions that are present in the governance of AI alignment, and provide an explicit analytical approach to examining them. We operationalize the framework through five analytical components, each with an accompanying "analyst recipe" that collectively identify the primary institutional logics and their internal relationships, external disruptions to existing social orders, and finally, how the structural risks of each institutional logic are mapped to a catalogue of sociotechnical harms. The proposed concept of structural transparency enables analysts to complement existing approached based on informational transparency with macro-level analyses that capture the institutional dynamics and consequences of decisions regarding AI alignment. 
\end{abstract}

\keywords{AI alignment, transparency, institutional logics, governance, sociotechnical harms}

\maketitle

\section{Introduction}
As generative AI diffuses across social and economic systems, the notion of AI alignment has expanded from a narrow technical view focused on incorporating principled “human values” to prevent misalignment with human intentions \cite{Jialignment2025}, to a broader examination of how designers construct value specifications and the social consequences of those choices \cite{gabriel2020artificial, korinek2022aligned}. This expansion of scope becomes apparent through two developments. First, alignment techniques increasingly try to accommodate pluralistic and heterogeneous values, reflected in social choice-inspired preference aggregation \cite{conitzer2024social}, constitutional AI \cite{huang2024collective}, and social feedback–based alignment \cite{liu2023training}. Second, because individuals, organizations, and institutions adapt to and reshape these systems, alignment is now understood as bi-directional; that is, AI systems should not only be evaluated in relation to established social and institutional structures, they should also be evaluated for how they reorganize those structures \cite{shen2025bidirectional}. These developments can be viewed from the lens of societal alignment, which emphasizes the need for better alignment between values that generative AI systems reproduce and those of \emph{diverse publics} \cite{ribeiro2018introducing}.

\begin{figure}[!t] 
    \centering 
    \includegraphics[width=.95\textwidth]{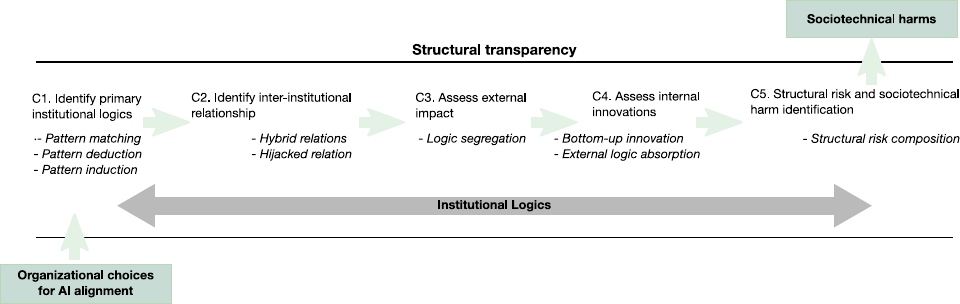} 
    
    \caption{Analytical components of Structural Transparency. Institutional logics inform the analytical components (C1–C5), which help analyze the sociotechnical impacts of organizational decisions for AI alignment.} 
    \label{fig:main_diag} 
\end{figure}

For responsible AI practices that deal with the social consequences of AI systems, this is a positive development. However, this expansion also requires new approaches to making the alignment processes transparent. Just as how the underlying models, data, and processes in pre-generative AI era act as the objects of transparency, generative AI adds new objects of inquiry. These consist of preference datasets, system prompts, normative principles, among other elements, which serve as constitutive inputs to AI alignment. Therefore, for the issue of transparency, to make the impacts at the societal level more visible, we need to complement the \textit{informational} approaches to transparency \cite{andrada2023varieties} with an approach that makes visible the underlying structural logic that guides organizational decisions around these constitutive inputs. Such approaches can, for example, highlight whether efficiency and productivity-based logic drives the underlying organizational rationale to alignment choices or result of a preemptive compliance from the pressures of the state. To that end, we develop the concept of \textit{structural transparency} (Fig. \ref{fig:main_diag}), which we define as the systematic analysis that makes visible the institutional structures and organizational decisions shaping the constitutive elements of alignment, and the ways aligned systems subsequently reshape those same structures once deployed. \par 

Before presenting an operationalisation of structural transparency, it is helpful to further articulate its motivating principles. The dominant paradigm for the analysis, evaluation, and development of AI systems is grounded in reduced-form methodological individualism \cite{hodgson2007meanings}, where the individual agent is the fundamental unit of inquiry. In the context of AI alignment, the normative goal of technical approaches is to develop systems whose behaviour coherently reproduces the idealised values of an aggregate of individuals \cite{gabriel2022challenge}. Even when there is a normative consideration that, like individuals, artificial agents should be socially embedded and that their reasoning be shaped by community norms, ethical, or relational obligations, these social aspects are ultimately reduced to the micro level. For example, Stańczak et al. \cite{stanczak2025societal} cast the problem of aligning AI systems with existing institutions in terms of a reduction to a principal–agent framework from microeconomic theory; Leibo et al. \cite{leibo2024theory} reduce the challenge of aligning model behaviour with diverse contextual norms to individual-level sanctions; altruism in socially aware AI systems is frequently reduced as internal reward shaping in reinforcement learning agents \cite{mckee2020social}. Although this methodological frame has technical utility as transparent mechanisms for incorporating social context into models, the reduction of all values to the individual (or an aggregation of individuals) is not adequate for understanding the social structural forces that shape the formation of values that alignment seeks to reproduce. Values are shaped by macro- and meso level entities, such as institutions, organizations, and communities, in which individuals are embedded \cite{friedland1991bringing}, and those social dynamics remain largely invisible if the analysis is anchored exclusively at the micro- (individual) level. Similarly, when dominant approaches to transparency \cite{lund2025standards}, which are based primarily on an investigation of the data and inner working of models through the explainable (xAI) frame \cite{minh2022explainable}, are applied to LLMs, they answer \textit{what} values the models reproduce \cite{scherrer2023evaluating, tao2024cultural, ji2024moralbench}, rather than \textit{how} those values are formed. Therefore, the motivation for developing structural transparency is to bridge this gap and develop a framework that makes explicit how alignment methods draw from (and potentially transform) organizations and institutions at the meso-level and macro-level. To that end, we make the following contributions. First, we introduce structural transparency as a framework for analyzing societal AI alignment that shifts the focus of transparency from model- and procedure-centric analyses to the institutional and organizational decisions that shape alignment and their downstream effects. Our approach draws on Institutional Logics \cite{thornton2012institutional} as the underlying theoretical foundation for this analytical framework. Second, we operationalize this framework through a five-component analytical procedure that supports a systematic analysis of transparency of alignment governance, including the identification of dominant and secondary institutional logics, their relationships, disruption pathways, internal organizational responses, and the resulting structural risks and sociotechnical harms. We further include a brief illustrative example to demonstrate the application of the framework in the context of an LLM-assisted tutoring system.

\section{Organizational decisions underpinning AI Alignment}
\label{sec:constitutive_inputs}

\small
\begin{table}[ht]
\centering
\begin{tabular}{p{2cm} p{2.8cm} p{4cm} p{5.2cm}}
\toprule
\textbf{Category} & \textbf{Techniques} & \textbf{Constitutive inputs} & \textbf{Organizational decisions} \\
\midrule

Learning from feedback  &
RLHF, SFT, RLAIF, Social Choice &
Pairwise preference comparisons \& rankings, SFT datasets, AI-generated synthetic datasets &
Annotator demographics and expertise; crowdsourcing vs. AI-based annotation; annotation guidelines; disagreement resolution protocols; quality-control mechanisms. \\

\midrule

Assurance &
Constitutional AI; Safety evaluations; System prompt engineering &
Principles/constitutions; evaluation rubrics; refusal policies; system prompts. &
Normative frameworks (legal and ethical principles) and guidelines for prescriptive alignment; benchmarks; documentation standards (model cards, safety reports).\\

\midrule
Governance &
Pre-deployment risk assessment; post-deployment monitoring and audits &
Risk and harm definitions; stakeholder feedback &
Risk and harm prioritization policies; accountability mechanisms; stakeholder inclusion; contractual arrangements.
\\

\bottomrule
\end{tabular}
\caption{Summary of constitutive inputs for AI Alignment and corresponding organizational decisions based on alignment techniques and categories in Ji et al. \cite{Jialignment2025}}
\label{tab:constitutive_inputs}
\end{table}
\normalsize

Organizations deploying gen-AI systems have control over the inputs that feed into the alignment process. Alignment methods such as RLHF, SFT, constitutional AI, and reinforcement learning with AI feedback rely on a range of artefacts, including preference datasets \cite{kaufmann2024survey}, annotation guidelines \cite{yuanuni} principle definitions, model-selection choices, benchmark criteria, and safety guardrails, that collectively determine the values that generative AI systems reproduce through their behaviour. Some organizations may have more control of the technical elements, and others may have more control over deployment choices. For example, a company deploying a consumer facing LLM-augmented device may have control over the stream of interaction data that feeds into fine-tuning, or adaptive alignment methods under distributional shifts \cite{thulasidasan2021effective}, while the underlying LLM model development organization has control over the technical alignment algorithm, both of which act as constitutive inputs to the overall aligned system. Ji et al. \cite{Jialignment2025} groups AI alignment into three categories (Table \ref{tab:constitutive_inputs}): (i) Technical alignment methods that learn from user or AI feedback; (ii) Assurance methods that concern system principles, prompts, and safety guardrails, which shape high-level behavioural constraints and normative commitments of the models; and (iii) Governance decisions around deployment choices, which determine what kinds of models (e.g., large vs. small, safety vs. capability trade-offs) are deployed,  risk and harm definitions for assurance \cite{anthropicUnderstandingAddressing}, and how stakeholder feedback is incorporated \cite{lim2025perceived}.

Human feedback-based learning methods form the backbone of technical alignment, such as RLHF (Reinforcement Learning with Human feedback) and SFT (Supervised Fine-Tuning). The procedure for gathering feedback typically follows one of two methods: prescriptive methods are jury-style evaluations where annotators assess outputs against a narrow set of principles, whereas, descriptive methods allow for subjective evaluations of the outputs based on broader principles \cite{kirk2023empty}. However, in both case, it is possible for subjectivity to enter the process through annotator biases \cite{peng2022investigations} or the normative frames through which annotators interpret tasks \cite{balagopalan2023judging}. Increasingly, in the RLAIF paradigm \cite{lee2023rlaif} (reinforcement learning with AI feedback, where AI systems act as evaluators), these processes are augmented with synthetic data generation using a higher-capacity model. All these methods involve a series of organizational decisions, for example, choices such as whether to select a prescriptive or a descriptive process to acquire feedback. Once the process is selected, decisions are made about the selection of annotators, including decisions about the target demographic background, domain expertise, crowdwork platforms, remuneration structures, and which stages of data set construction should be automated or partially automated using AI. Organizations also set quality-control criteria that govern the reliability of annotation and disagreement resolution. All such decisions form the category of assurance and governance of AI alignment, and the main objects of inquiry, with reference to structural transparency, are these decisions within AI alignment.

\begin{figure}[!t] 
    \centering 
    \includegraphics[width=0.9\textwidth]{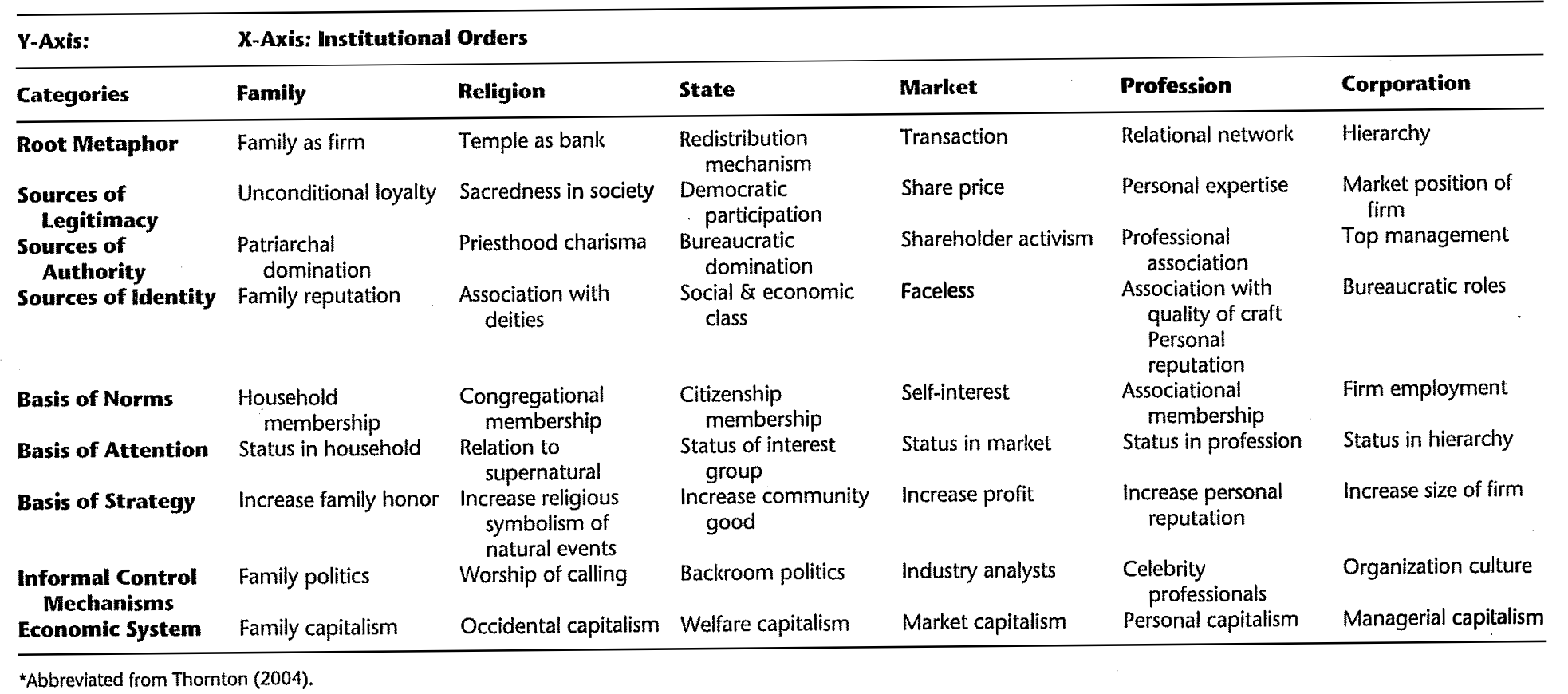} 
    
    \caption{'Ideal types' of institutional orders (reproduced from Thornton and Ocassio 2012 \cite{thornton2012institutional}.)} 
    \label{fig:inter_inst_system} 
\end{figure}

\section{Institutional Logics as the basis for structural transparency}
Our reference to the \textit{values} reproduced by generative AI systems is grounded in the notion of value-sensitive design and the recognition that values in sociotechnical systems include the institutional norms and reasoning patterns of actors that shape technical design choices \cite{van2020embedding}. We take the analysis of these values to the structural level with the aim of elucidating the aetiology of values as they become embedded in organizational decisions. To this end, we draw on the theory of \emph{Institutional Logics} \cite{thornton2012institutional}, a dominant framework in organizational studies for analyzing how socially constructed belief systems and organizing principles shape actors’ understandings of rationality and appropriate action. In this section, we provide a brief background on the theory.\par
As a definition, Institutional Logics refers to the \textit{socially constructed historical patterns of cultural symbols and material practices—including assumptions, values, and beliefs—through which individuals and organizations make sense of their activity, organise time and space, and reproduce their lives and experiences} (Thornton \& Ocasio, 2008 \cite{thornton2008institutional}). The theory puts into work the idea that individual rationality and sensemaking are contingent on the institutional structures that shape everyday experience of the decision-making entities. At the societal level, institutional logics constitute a set of `ideal types' of institutional orders that define modern society--the family, community, religion, state, market, profession and corporation. Although Institutional Logics draw from the Weberian ideas of values \cite{weber1995religious} as theoretical antecedents, the theory goes beyond that foundation by suggesting that institutional orders provide distinct cognitive schemas of meaning and legitimacy that collectively shape the underlying social construct of \emph{rationality}. Each institutional order (Fig. \ref{fig:inter_inst_system}) provides distinct epistemic grounding of a `logic' which are defined by their categorical elements (Y axis in Fig. \ref{fig:inter_inst_system}) consisting of distinct norms (what behaviour is considered appropriate and acceptable), legitimacy (on what grounds are the organization's decisions justified), authority (who is entititled to make binding decisions), identity (who the organization serves), attention (what information, concerns, risks, do organizational actors prioritize) that collectively guide strategic behaviour within organizations and institutions. At the field level, networks of organizations and communities operate by embodying and reproducing the logics of (often conflicting) institutional orders they draw upon. At the micro level, individuals embedded within these fields reproduce shared understandings of what counts as "rational" and "appropriate", while also contributing to institutional change by influencing the very entities that structure their behaviour.  \par
Beyond the ideal types of institutional orders (and logics) shown in Fig.~\ref{fig:inter_inst_system}, institutional logics also support the identification of hybrid categories, sub-categories, and field-level operationalizations that reflect distinct patterns of institutional reasoning. For example, academic logic may be understood as a derivative of professional logics as operationalized in the field of higher education \cite{cai2022institutional}, while community-based logics manifest as communities of place, users, and firms \cite{georgiou2023community}. Similarly, the field of healthcare can be guided through field-specific science- and care-based logics \cite{fincham2015three}. In the context of our proposed framework, institutional logics provide the analytical lens through which alignment-relevant decisions are conceptualized as outcomes of institutional reasoning, which moves away from viewing such decisions as a result of neutral or purely objective rational decision-making.

\section{Structural transparency}

Fig. \ref{fig:main_diag} shows the high-level relations between the core components of structural transparency. The main core of structural transparency is built around a set of five analytical components that present the set of institutional questions necessary to make institutional dynamics visible. In the language of bi-directional AI alignment \cite{shen2025bidirectional}, the first two components (C1 and C2) focus on forward alignment by identifying the institutional logics that drive organizational decisions in constructing constitutive inputs. The next two questions (C3 and C4) turns to backward alignment, highlighting how these alignment choices may disrupt existing institutional orders that do not directly participate in the alignment process. Finally, C5 translates these institutional dynamics onto structural risk categories and help analyze how these dynamic manifests into potential sociotechnical harms. We develop and present each component as “analyst recipes,” which analysts can use to operationalise the structural transparency framework.

\subsection*{C1: Identifying primary and secondary institutional logics guiding alignment}
This question identifies which set of institutional logics, such as market, professional, corporate, community, or state, implicitly guide the construction of constitutive alignment inputs, like preference datasets, annotation protocols, and constitutional principles. We present a recipe for this analytical component using Reay and Jones' \cite{reay2016qualitatively} pattern-based approach to identifying the logics. 

\subsubsection*{Pattern matching, deduction, and induction.}  The first approach, based on pattern \textit{matching}, involves constructing examples of behaviour and practices that would constitute ideal types of institutional logic based on how each categorical element of the corresponding logic would manifest in the organizational decisions of alignment. The analysis begins by hypothesising the set of institutional orders relevant to the alignment process. Using the categorical elements associated with each order (as illustrated in the Y-axis of Fig.~\ref{fig:inter_inst_system}), the analyst systematically examines how decision rights and procedural norms are organised within the alignment pipeline. For example, does the organization treat community-level judgments as legitimate inputs by strictly centering a community logic through social-choice-based preference aggregation, or are preference construction services outsourced to a market intermediary operating under market-based logics \cite{clark2019regulatory}? Each such categorical element on the institutional-logics Y-axis is then matched to the order whose characteristics most closely describe how the alignment process is structured. The second approach, through pattern deduction, involves quantitative analysis methods such as topic modelling and word frequency count on sources of textual data that contain or reflect organizational decisions. Common sources of this data include organizational documents (e.g., policy statements, strategy documents), records of decision-making processes, media releases, and public or internal communications by organizational leaders \cite{madsen2025podcasting}. In contrast to matching based on the categorical elements of an institutional order, the researcher would have \textit{a priori} conceptualisations of the words and phrases that represent each logic, such as "return on investment" as corporate logic, "fairness metrics" as professional logics, or "value for money" as market logic. They would then quantify these concepts using word-frequency-based or other NLP methods \cite{patil2023survey} to establish which set of logics dominate the sources \cite{reay2016qualitatively}. Finally, the third approach is pattern induction, which relies on a grounded theory-based approach using bottom-up evaluation through interviews, direct observations, and other ethnographic methodologies. Instead of a \textit{a priori} conceptualization of the constructs of each logic, the analyst lets the sources naturally build an ontology for each logic using interpretive methods that directly reference the original sources \cite{rowlands2005grounded}.

\subsection*{C2: Identifying relationship between primary and secondary logics}
While the identification of primary logics (C1) establishes which institutional orders are present, this analytical component examines the nature of relationship \emph{within} the set of primary logics. In particular, this component distinguishing cases in which a secondary logic meaningfully is integrate into organizational decisions from cases in which they are invoked cursorily as means to a strategic end. To do so, we focus on two distinct relational forms through which multiple logics may be embedded in organizational practice: hybrid logics and hijacked logics. 
A hybrid logic arises when organizational decisions are substantively guided by more than one institutional logic, such that each logic meaningfully informs decision-making criteria, organizational practices, and accountability structures. For example, when an organization adopts a participatory approach to AI alignment \citep{delgado2023participatory}, affected stakeholder communities are granted a substantive role in shaping design and governance choices. In such cases, alignment decisions are situated simultaneously within a community logic and a safety logic. When there is genuine adherence to the principles and commitments of the secondary logic, rather than merely invoking its language, this configuration can be characterized as a true hybrid logic. By contrast, a hijacked logic refers to the strategic invocation of a secondary logic primarily for persuasive purposes, rather than as a genuine guide to organizational action. In this case, organizations "speak the language" of a secondary logic while continuing to organize decisions according to the dominant logic. The secondary logic functions instrumentally, as a rhetorical or symbolic resource towards advancing organizational objectives, rather than as a source of constraint or guidance in practice \cite{mcpherson2013logics}. 

\subsubsection*{Distinguishing hybrid and hijacked logics.} Distinguishing between hybrid and hijacked logics is often analytically challenging, as both may involve explicit references to multiple logics and it is hard to infer intent from observed organizational actions. To address this, we build upon McPherson and Sauder \cite{mcpherson2013logics} and propose the following operationalizable analytical recipe.
First, analysts must identify and exclude contexts in which constraints prohibit the meaningful application of a secondary logic within a given decision frame. For example, it may be implausible to expect the application of a community logic in safety analysis to prevent inadvertent disclosure of information critical to national security, or the moderation of content that is legislatively prohibited within a jurisdiction. In such cases, the absence of meaningful integration of secondary logic does not constitute an informative signal on hybridisation or hijacking. McPherson categorizes such limitations as procedural constraints, including regulatory requirements and standard operating procedures that explicitly preclude the invocation of certain logics in specific contexts. Second, analysts must eliminate decision contexts characterized by definitional constraints, where the application of a secondary logic is implausible by definition. These are cases in which the scope of a logic render its application incoherent within a particular organizational decision, independent of strategic intent. Following the previous example, applying a community logic to low-level numerical hyperparameter tuning is a mismatch by definition since it is not the level of abstraction and context in which a broad-based community participation or deliberation can meaningfully operate. 

Once contexts subject to procedural and definitional constraints are excluded, analysts can assess whether the secondary logic is meaningfully integrated into the organization’s principles, practices and structures, or whether it is invoked only superficially. A hybrid logic manifests itself as a durable organizational arrangement in which multiple logics are jointly sustained over time. For example, Skelcher and Smite \cite{skelcher2015theorizing} document the case of a restaurant that employs disadvantaged youth under a government-supported programme. Such structures are examples of hybrid logics that integrate a market logic of commercial viability with a community logic of social inclusion and a state logic through public contracting. Contrast this to to cases in which organizations invoke ethical principles instrumentally, for example, by washing ethics \citep{bietti2020ethics} or superficially to achieve the goal of entering new markets \citep{tan2011mnc}, without substantively reorganising their operations around an ethics-based logic \citep{hanson2009ethical}. We use three indicators to assess the relationship:

The first indicator is temporal durability: does the secondary logic persist across an extended period or is it applied only episodically for specific purposes? The second indicator is organizational integration: Does secondary logic shape core practices such as evaluation, accountability, and governance, or is it confined to a narrow set of isolated decisions? The third  concerns conflict resolution; when tensions arise between the logics, does the organization maintain coherent mechanisms, such as role separation and responsibility to manage these conflicts? The presence of credible and sustained conflict-management processes indicates that the organization has established a process in place to manage the conflicts arising from a hybrid logic. Conversely, the absence of such mechanisms could suggest a hijacked relationship. For example, ethics-based logic is inherently in tension with the primary corporate or market logic that governs for-profit firms. If an organization offers no credible account of how such conflicts are resolved in practice, the invocation of ethics might not be credible evidence of a true hybrid configuration.

\subsection*{C3: Assessing external disruptions}
The primary objective of this component is to assess the effects of the logics that shape AI alignment on existing social structures and processes. We do so by analysing structural impact at the level of a target field (or domain), a network of institutions and organizations whose interactions form the shared rules of practice in a domain. For example, the field of higher education comprises universities, accreditation bodies, and funding agencies; the field of healthcare comprises hospitals, insurers, professional colleges, and state agencies; and the field of AI safety and alignment comprises frontier labs, research labs, regulators, ethicists, and standards bodies. The earlier analytical components (C1–C2) establish the logics that shape alignment, while this component shifts the unit of analysis outward by asking how the institutional logics embedded in alignment practices, when reproduced through the outputs and operating assumptions of an aligned system, exert pressures on external fields that are governed by their own logics. This question adopts a \emph{structural} perspective on the broader concerns about the disruption that gen-AI systems can bring about because it interprets the disruption as a change of the rationality (i.e., what counts as success and how it is measured, what counts as acceptable justification) of how the target field operates, rather that looking solely through a lens of technology adoption \cite{puavualoaia2023artificial}. When an `aligned' AI system is introduced into a target field, its behaviour and affordances carry traces of the dominant logic that governed its design and alignment. For example, if an aligned system’s output replicates an efficiency-orientated corporate logic, e.g., prioritising throughput, standardisation, and managerial control, because that is what guided the annotation principles in the fine-tuning dataset, then when that system is adopted into an external field such as public management or higher education, it will activate institutional pressures on the logics of the target field. The central analysis in C3 is therefore to make these impacts transparent and analyse how insulated or permeable are the boundary of the logics of the target external field. To operationalize this, we use the theory of logic "segregation" \cite{thornton2012institutional, skelcher2015theorizing}. Segregation refers to the degree of separation between institutional orders. Higher segregation implies that logics occupy distinct spheres of influence, with fewer opportunities for cross-order encroachment; lower segregation implies that logics are already penetrating each other, increasing the likelihood that the categories of legitimacy and rationality of one order can reshape the categories of another. On this basis, we propose the following analytical recipe that isolate a plausible disruption pathway:

\textit{Step 1: Identify the target field's institutional logics.} The analyst begins by selecting an external field whose practices are plausibly affected by adoption of the aligned system (e.g., higher education). The analyst then identifies the primary and secondary institutional orders that structure that field. For example, the field of higher education can be assessed as primarily embedded in academic logic, with market logic operating as secondary logic in a hybrid relationship (e.g., competition for student recruitment and rankings coexisting with academic norms and academic freedom \cite{grossi2020impact}).

\textit{Step 2: Locate overlap as a potential disruption pathway.} From C1–C2, the analyst carries forward the dominant and secondary logics governing the alignment process. Taking that set along with the ones generated in the above step, the analyst finds the intersection of the two. This intersecting logic isolates a specific pathway for institutional pressure. For example, the alignment field may be assessed as primarily embedded in corporate and market logics, with professional and safety logics invoked as secondary logics that constrain certain classes of outputs. Continuing the example of higher education, market logic appears in both alignment decisions and in higher education, and therefore the intersecting logic. Due to this intersection, the behaviour of the AI system can likely amplify the rationales based on market-logic that already have influence in the target field. This step therefore identifies this disruptive pathway in the form of market-based logic of how AI might impact higher education.

\textit{Step 3: Assess the disruption through degree of segregation;} Intersection alone might not imply disruption. The analyst needs to evaluate whether the target field has established conflict-resolution or separation mechanisms that isolate areas of control between its hybrid logics. In higher education, such mechanisms might include governance structures that neatly isolate arenas of academic decision-making from revenue imperatives (e.g., strong faculty authority over curriculum, tenure protections). These are segregation mechanisms that limit where the disrupting market logic can be legitimately applied. If such separation is absent or fuzzy, the analyst can narrow down how adoption of the system will generate disruptive pressures, and which logic will plausibly supply the justification for those pressures. In the example, if academic governance lacks credible mechanisms to contain market imperatives, then adoption pressures surrounding the aligned system will tend to be rationalizable in market terms; for example, AI adoption framed as necessary for competitive positioning in student recruitment and ranking performance. 

The result of the above analysis identifies the underlying structural forces that result in changes in practice from the adoption of an aligned system. The analytical output from this component identifies, first, which institutional orders from C1-C2 create a plausible disruption pathway in external institutional structures and, second, under what conditions and what form that pathway is likely to destabilise existing practices of the external field.

\subsection*{C4: Assessing internal disruptions and innovations.}

This component analyses how pressures from external institutional orders are incorporated into an organization’s internal logic when making alignment decisions. This component can be thought of as a precursor to logics being integrated into the decision-making process of the sort identified by C1. The focus of this component is on identifying the inbound pressures, that is, how external logic affects the organizational reasoning that governs alignment decisions. We develop the following analytical recipe to make this dynamic transparent.\\
\subsubsection*{Response to external logics' pressures.} The analyst first starts with a focal field for analysis that has agency to make organisational decisions around some part of the constitutive input to alignment (c.f. Sec \ref{sec:constitutive_inputs}). This may be a set of organizations involved in developing aligned AI systems (e.g., frontier or research laboratories) or organizations responsible for integration and deployment in a specific domain (e.g., public agencies, universities, or healthcare providers). Next, the analyst identifies the external institutional orders that exert pressure on the organization’s alignment decisions. These pressures may originate from state logic (e.g., regulatory or national-interest pressures on model behaviour), community logic (e.g., demands for participatory governance or considerations for affected groups), market logic (e.g., competitive pressures to adopt or accelerate deployment), or professional logic (e.g., standards of acceptable practice). Such pressures arise from external orders that are directly or indirectly affected by the aligned system and often act as precursors to hybrid institutional reasoning (c.f. C2) if they are taken up within organizational processes. Previous organizational research has shown, for example, that community-based logics can drive bottom-up innovation by reshaping organizational practices \cite{lee2015filtering}. Similarly, research on public-sector governance has shown how pressures from market- and efficiency-orientated logics have driven the adoption of digital technologies in public service delivery, reshaping organizational decision-making practices in domains such as e-government and digital public administration \cite{vickers2017public}. Finally, through qualitative analysis of organizational artefacts (e.g., interviews, press releases, statements), the analyst evaluates whether and how these external pressures are absorbed. The nature of analysis is anchored to changes in categorical elements (Y axis Fig. \ref{fig:inter_inst_system}) of institutional logic by which the decisions are shaped (which is already identified in C1); and when external pressures are found to be substantively incorporated, they appear as a hybrid logic appearing in the analysis in C2.
\small
\begin{table*}[!t]
\centering
\small
\begin{tabular}{p{2.5cm} p{3.5cm} p{7.5cm}}
\toprule
\textbf{Institutional Logic} & \textbf{Structural Risk Category} & \textbf{Sociotechnical Harms (Shelby et al. 2023) \cite{shelby2023sociotechnical}} \\
\midrule
\multirow{2}{*}{State}& Surveillance \cite{hu2017national} & Interpersonal harms: privacy violations; Representational harms: loss of autonomy \\
& Information control \cite{chen2017information} 
& Information harms: misinformation, disinformation, malinformation \\
\midrule

\multirow{3}{*}{Profession} 
& Technocratic gatekeeping \cite{bertsou2020technocratic} 
& Interpersonal harms: Loss of Agency \\
& Regulatory capture \cite{makkai1992and}
& Quality of Service harms: Alienation, Service/benefit loss \\
\midrule

\multirow{3}{*}{Market} 
& Market manipulation / asymmetry \cite{grossman1980impossibility}
& Allocative harms: opportunity loss; Quality-of-service harms: increased labour \\
& Market failure \cite{randall1983problem}
& Socioeconomic system harms (e.g., distorted market access, increased inequality, labour insecurity) \\
\midrule

\multirow{3}{*}{Corporation} 
& Shareholder capitalism \cite{engelen2002corporate}
& Socioeconomic system harms (e.g., distorted market access, increased inequality, labour insecurity) \\
& Corporate surveillance \cite{christl2017corporate} 
& Interpersonal harms: privacy violations; Representational harms: loss of autonomy \\
\midrule

\multirow{3}{*}{Religion} 
& Religious persecution \cite{rifat2024politics}
& Representational harms: stereotyping, misrecognition \\
& 
& Interpersonal harms: harassment, violence \\
\midrule

\multirow{3}{*}{Family} 
& Nepotism \cite{szakonyi2019princelings}
& Allocative harms: opportunity loss; Quality-of-service harms: increased labour \\
& Coercive control \cite{reeves2025dangerous}
& Interpersonal harms: technology-facilitated violence, privacy violations \\
\midrule

\multirow{3}{*}{Community} 
& Inter-community conflict \cite{unver2024artificial}
& Interpersonal harms: harassment, violence \\
& Epistemic bubbles \cite{nguyen2020echo}
& Information harms: misinformation/disinformation; Representational harms: stereotyping \\
\bottomrule
\end{tabular}
\caption{Mapping institutional logics to structural risk categories and sociotechnical harms from Shelby et al. (2023) \cite{shelby2023sociotechnical}.}
\label{tab:strctural_risks}
\end{table*}
\normalsize
\subsection*{C5: Identifying structural risks and sociotechnical harms}
This component makes transparent the pathways through which the organizational decisions that underpin alignment processes manifest themselves as structural risks and downstream sociotechnical harms. Each institutional order carries characteristic benefits and structural risks. These benefits and risks define the societal trade-offs that are implicitly manifest when alignment is governed by a corresponding logic. For example, organizations guided by market logic may realize the efficiency benefit of adopting the optimal alignment technique for the compute budget, while also carrying the structural risk of market failure, which can manifest as externalities of social and environmental costs. C5 connects these institutional dynamics to material outcomes by mapping logics to structural risks and then structural risks to concrete sociotechnical harms. This component is geared to answer two questions: (1) What sociotechnical harms are plausibly exacerbated by the primary and secondary logics that govern alignment? (2) In a specific deployment context, what harms are actually materialized or mitigated by the composite logic arrangement? We present the following recipe to answer the above questions:\\
\subsubsection*{Structural risk composition.} 
In the first step, the analyst compiles the "logic set" that will be evaluated for sociotechnical harms by carrying forward the outputs from C1-4: the primary logics identified in C1, their relationship from C2, the external logics implicated by the deployment pathway in C3, and any novel hybrid logic identified in C4. 
For example, if C1–C2 identify corporate + market as dominant logics in alignment, with professional + safety as secondary logics, and C3 identifies higher education as the target field embedded in academic + market logics, then the candidate logic set is {corporate, market, professional, safety, academic}.

The second step maps each logic in the set to baseline structural risk and benefit categories. Each institutional order in the logic set is mapped to a baseline bundle of structural risks and benefits that it characteristically carries. Table \ref{tab:strctural_risks} provides the risk/benefit categories that the analyst can draw from (e.g., surveillance and authoritarian control for state logic; shareholder capitalism and monopolistic practices for corporate logic; externalities and commodification pressures for market logic; gatekeeping and exclusion for professional logic; solidarity and care with risks of paternalism for family/community variants; academic logic with benefits of epistemic standards and risks of credentialism or exclusion).

The third step translates the structural risks into a candidate list of sociotechnical harms that can plausibly arise at the interpersonal, organizational, and societal levels. We use the taxonomy of Shelby et al. to construct this mapping which takes as input the structural risk of a given institutional logic and identifies its manifestation in a predictable family of harm. For example, a dominance of state logic’s structural risk of surveillance and authoritarian control maps to interpersonal harms such as privacy violations and loss of autonomy; societal harms such as chilling effects and political repression. Similarly, dominance of corporate logic’s structural risk of shareholder capitalism and monopolistic practices maps to societal harms of increased inequality, distorted market access, and exclusionary control over resources. The output is a 'candidate' set of harms that are structurally plausible given the institutional ordering of the system.

Finally, in the fourth step, the analyst generates the composite effect of the structural risks in the context of the deployment of the system. Having identified a candidate sociotechnical harm set, the analyst evaluates which harms are likely to materialize under a specific deployment case and whether composition of the logics neutralizes or exacerbates them. This step is operationalized through a risk composition analysis procedure using two inputs: (i) the structural risk/benefit matrix for the logics compiled in Step 1; and (ii) information artefacts that characterize how the system is actually deployed and governed (e.g., agreement, contracts, press releases, policy documents, usage policies, staffing decisions, and accountability mechanisms). Based on these inputs, the procedure consists of an analysis of how deployment conditions activate or constrain the structural risks associated with the set of institutional logics. Structural risks are materially translated as sociotechnical harms when activated by deployment choices; conversely, harms are mitigated when deployment embeds constraints by operationalizing the benefits and accountability structures of other logics in the set. For example, consider the illustrative example (discussed in Sec. \ref{sec:illustrative_example}) of a university’s deployment of an LLM-based tutoring service provided by an external vendor fine-tuned based on past data from online course forums. If the primary logic identified is corporate logic, it introduces a structural risk of labour precariousness, which can manifest itself as the sociotechnical harm of reduced on-campus employment for graduate students and diminished training opportunities. The composition analysis asks whether the university’s deployment arrangement treats tutors as substitutes (replacement) or complements (augmentation). If the service substitutes graduate tutoring with weak academic oversight, then the structural risk of corporate logic is activated, and the damage is exacerbated since the precarity of employment materialises as a result of the deployment. If the service is configured as a complement (e.g., tutors retain agency over when and how the tool is used; the system is integrated as an apprenticeship support with academic governance), then the academic logic provides a constraint that can mitigate the structural risk from corporate logic. Finally, the output of this process is the set of sociotechnical harms that are residual after the composition step. These are the harms that are not neutralized by any countervailing logics.

\section{Illustrative example}
\label{sec:illustrative_example}

We use the following stylised case as a brief illustration of the structural transparency framework. We note that this analysis is only illustrative using a hypothetical case, and a detailed operationalisation and analysis of real-world deployment case study would warrant a separate contribution.

\begin{quote}
\textit{In order to expand instructional support capacity in large undergraduate courses, a public research university deploys a third-party large language model (LLM)–based tutoring system. The system is deployed through an external vendor under a long-term service contract, and the vendor develops and maintains a general-purpose LLM, which it has aligned for educational use through supervised fine-tuning and reinforcement learning from human and AI feedback based on past student support data held by the university. The vendor retains control over core alignment decisions, including the usage of preference datasets, safety principles, evaluation benchmarks, and the option of future pricing changes. Limited reconfigurability is provided to the university through system prompts and access control mechanisms. The university has evaluated the system using safety reports provided by the vendor and basic testing against internal metrics of instructional quality. Within the university, responsibility for adoption and oversight is distributed through a committee that includes central administration, technology services, and faculty leadership. The university has integrated the system into the existing learning management framework and avails configuration choices to align the system’s interaction tone through system prompts. The system logs interactions for monitoring and quality assurance, and students are directed to raise usage issues and technical support tickets directly with the vendor through a centralized reporting interface. Although official communications frame the system as complementary to human tutoring, internal planning documents reference anticipated efficiency gains and potential savings in graduate student labour. The graduate student association has publicly opposed the use of historical student–tutor interaction data for model fine-tuning, citing concerns over consent of the use of past intellectual output and labour displacement.}
\end{quote}

\noindent (\textbf{C1.}) C1 focusses on the identification of primary institutional logics that guide the alignment governance of the system. In this case, it includes both the decision to deploy the AI tutoring system and the specific design choices associated with that deployment. Using a \textit{pattern-deduction} approach, we identify the closest matching institutional orders characterized by the categorical elements that define a particular institutional logics. In that context, although universities are situated within academic research community-based logic, the practice of tutoring is primarily driven by professional pedagogical principles. This aligns the broader university as a mix of community and professional logics \cite{cai2022institutional}; whereas, the specific use-case of the deployment aligning closer to the latter. The source of legitimacy (what makes the organizational choice of AI deployment justifiable) draws on market-based imperatives to remain competitive within the higher-education field. Similarly, the basis for attention (what the organisation prioritises in its decision-making) is also embedded in the market logic. Based on this brief analysis, we can propose that the professional and market logics constitute the primary institutional logics that govern the organizational decisions.\par

\noindent (\textbf{C2.}) For C2, we characterize the interaction between the professional and market logics as a hybrid rather than a hijacked relation. The long-term nature of the contract provides this relationship with temporal durability, satisfying the first criterion for a hybrid inter-logic relationship. When conflicts arise between the two logics, resolution is distributed since technical issues are delegated to the vendor, while the university committee retains the final say over deployment decisions subject to contractual constraints. The two logics are highly integrated in the organization decisions as observed by the market-driven imperative for the deployment. Therefore, based on the three criteria for C2, we can classify the relationship as a hybrid. \par

\noindent (\textbf{C3.}) C3 identifies the pathways of disruption to a field; in this case, the field of higher education. Higher education is traditionally rooted in a hybrid mix of community-based academic logics and professional logics that maintain quality standards \cite{kallio2021institutional}. The interaction between the deployment’s logics (market + professional) and the university’s existing primary logics identifies the market logic as a potential disrupting force \emph{on} academic logic. Such a disruption can manifest as an increase in market-oriented views within adjacent arenas of university operations, such as academic research; for example, through a push of initiatives where the university provides research services in an open marketplace \cite{bingham2011open}, thereby disrupting existing practices of research. Unless an organizational process is in place to segregate the impact of market-based logic from academic operations, a deployment that may at first glance appear unrelated can create a future pathway for this structural disruption.\par

\noindent (\textbf{C4.}) This component answers how the primary logics that guides organizational decision (professional+market, in this context) incorporate pressures from external institutional logics that lie outside those primary logics. In this case, community-based logic exerts pressure through graduate student associations, which oppose the use of their intellectual output derived from past student support interactions for fine-tuning models. There is also opposition on the ground that the system ultimately substitute for graduate student labour. An analysis under C4 should therefore draw on interviews with university administrators and graduate students to assess whether the university adopts innovative governance practices that meaningfully integrate these pressures, which, if incorporated, would appear as elements of the primary institutional logics. In the absence of such innovation, C4 identifies this conflict as a persistent inward pressure that challenges organizational decisions around the deployment.\par

\noindent (\textbf{C5.}) This component identifies the sociotechnical harm emerging out of the key institutional logics of the deployment context. The set of institutional logics in focus is comprised of market, professional, and academic community logics. The structural risk mapping (Table \ref{tab:strctural_risks}) identifies that market logic carries risks of market failure, which map to sociotechnical harms of graduate labour insecurity and the compromise of pedagogical standards (quality-of-service). Since the oversight committee includes faculty representation, the professional logic provided by the committee can mitigate the degradation of pedagogical standards. However, harms related to labour disruption and data-ownership rights remain unmitigated. Although community-based logic can potentially offer mitigation through participatory governance and improved data-use negotiations, these sociotechnical harms persist because the organizational decision fails to incorporate that logic in an innovative way (as identified in the C4).

\section{Related work}
\noindent \textit{Approaches to transparency:} The overarching goal of transparency is to render the mechanisms and consequences of AI systems legible to affected stakeholders \cite{larsson2020transparency}. There are a diverse set of perspectives to the notion of AI transparency, notably, informational, material, and procedural transparency \cite{diakopoulos2020197, andrada2023varieties}. Informational transparency, the most prominent strand, makes visible the algorithms, data provenance, safety evaluations, fairness assessments, and contextual appropriateness of system use \cite{mitchell2019model, larsson2020transparency}. When adapted to generative AI and post-training alignment pipelines, such transparency would involve systematically documenting the constitutive elements of alignment methods through an enquiry into the details of the datasets used for alignment methods, how annotator judgments are aggregated, and how high-level principles in constitutional AI are formulated and interpreted. Material transparency concerns the infrastructures and environmental impacts from the practical implementation of alignment \cite{andrada2023varieties}. Procedural transparency examines the governance mechanisms surrounding these practices, including elements such as, oversight structures, auditing protocols, and redress mechanisms \cite{buijsman2024transparency}. Although extending these transparency frameworks to generative AI alignment introduces novel challenges, such as translating high-level principles into operational and auditable frameworks, these issues remain distinct from the structural concerns raised by the societal expansion of the AI alignment problem. In this context, structural transparency can be considered a specific instance of \emph{value-based} transparency \cite{buijsman2024transparency}, grounded in value-sensitive design \cite{friedman2013value} and the perspective that the values driving designers are embedded within system design. Structural transparency advances this perspective by making explicit how those values are constructed from underlying social and institutional structures. We advance this approach by operationalizing key elements of value-based transparency developed by Buijsman \cite{buijsman2024transparency}, \emph{vis-à-vis} the identification and conceptual specification of different values within a relevant socio-technical system, and the elucidation of how value conflicts are resolved.\\
\noindent \textit{Transparency of AI alignment:} The notion of \emph{values} are central to AI alignment and there is extensive literature from a normative frame of proposals of how these values \emph{should} be constructed. Notable approaches are based on participatory methods \cite{bergman2024stela, kellmeyer2024beyond, birhane2022power}, norm-based approaches with contextual social reasoning \cite{leibo2024theory}, and social choice \cite{conitzer2024social}. Our work is situated within a descriptive frame that proposes a framework of transparency into how values \emph{are} constructed. Existing literature within that descriptive frame, use interpretive methodologies as the main methodological grounding \cite{law20161}. For example, Deng et al. \cite{deng2023investigating} use inductive thematic analysis to understand how industry practitioners facilitate cross-functional collaboration for AI fairness by piggybacking on institutional procedures, such as, impact assessment towards furthering fairness as a value in AI systems. Chen et al. uses ethnographic data to study state-driven mobilization of moral values of AI in context of China \cite{chen2025moral}. Rudschies et al. \cite{rudschies2020value} use content analysis of ethics guidelines documents to build an actor-centric approach to highlight the divergences in priorities for ethical principles across public, expert, and private actors in the AI ethics debate. Compared to these approaches, our framework is institution-centric since it methodologically integrates the macro-level analysis provided by institutional logics into a transparency framework. Second, compared to the dominant interpretive approaches to descriptive analyses of values in AI systems, our framework shifts the unit of analysis from individual sensemaking to organizational decisions. These decisions are compatible with an analysis using a rational choice lens, while still allowing for subjectivity through the institutional logics that shape actors’ understandings of rationality.

\section{Discussion and Conclusion}

The main contribution of the paper is a framework for the analysis of organizational decisions around the alignment of generative AI systems. In contrast to existing non-technical treatments of AI alignment that have overwhelmingly focused on the identification of values that generative AI systems should reproduce, this paper brings into scope the organizational decisions that drive that process. Societal governance of disruptive and innovative technologies often faces a dilemma that public institutions must address the social impacts of innovative technologies with urgency, whereas clarity and certainty regarding the societal effects of such technologies often emerge only decades later. This dilemma, identified by Collingridge \cite{collingridge1980social, ribeiro2018introducing}, is unsurprisingly also present in the context of generative AI systems, where much of the governance of alignment is left to private organizations. Although this dilemma highlights an inherent challenge, there is a growing recognition of the urgent need for accountability for these organizational decisions. To that end, our paper advances the conversation in two ways. First, transparency (construed broadly) is a necessary condition for accountability, and the proposed framework advances transparency by identifying (through components 1, 2, and 4) the underlying logics and rationalities that drive organizational decisions around alignment. Second, the framework adds clarity to the societal impacts of the disruptive technology of generative AI systems (through components 3 and 5) and therefore takes a step toward easing the dilemma. Although the focus of the paper has been on organizational decisions, the framework can also be adapted to analyze AI policy. This adaptation would require minimal changes to components 1, 2, and 5, since the identification of underlying logics, their internal relationships, and sociotechnical risks remains relevant when the object of analysis shifts from organizational decisions to AI policy. Components 3 and 4 would require reframing in the policy context, as these components are oriented toward specific deployment settings, whereas organizational or institutional AI policy is, by definition, broader than any particular deployment context.\par
In line with the scope and focus of this paper, we identify two immediate directions for future work. First, the illustrative example presented here is intended only as an initial demonstration of the framework and should be extended to a real-world case study that places further emphasis on the empirical parts of the analytical components. Second, future work should more explicitly examine how macro-level institutional logics are instantiated at the micro-level reasoning and (bounded) rationality of individual actors, thereby strengthening the connection between social and institutional structures and individual decision-making.

\newpage
\section*{Endmatter}
\noindent \textit{Generative AI usage statement}: In this work, Generative AI was not used to generate text. Its use was restricted to minor editorial assistance with formatting, grammar, and fluency. 
\bibliographystyle{ACM-Reference-Format}
\bibliography{sample-base}


\end{document}